\documentstyle[11pt,aaspp4]{article}

\def\la{\mathrel{\hbox{\rlap{\hbox{\lower4pt\hbox{$\sim$}}}\hbox{$<$}}}}
\def\ga{\mathrel{\hbox{\rlap{\hbox{\lower4pt\hbox{$\sim$}}}\hbox{$>$}}}}
\def\lesssim{\mathrel{\hbox{\rlap{\hbox{\lower4pt\hbox{$\sim$}}}\hbox{$<$}}}}
\def\etal{et al.\,\,}

\slugcomment{Accepted to {\it The Astrophysical Journal (Letters)}}

\begin{document}

\title{GRB 970228 Revisited:  Evidence for a Supernova in the Light Curve and Late Spectral Energy Distribution of the Afterglow}

\author{Daniel E. Reichart}

\affil{Department of Astronomy and Astrophysics, University of Chicago, 5640 South Ellis Avenue, Chicago, IL 60637}

\begin{abstract}
At the time of its discovery, the optical and X-ray afterglow of GRB 970228 appeared to be a ringing endorsement of the previously untried relativistic fireball model of gamma-ray burst (GRB) afterglows, but now that nearly a dozen optical afterglows to GRBs have been observed, the wavering light curve and reddening spectrum of this afterglow make it perhaps the most difficult of the observed afterglows to reconcile with the fireball model.  
In this Letter, we argue that this afterglow's unusual temporal and spectral properties can be attributed to a supernova that overtook the light curve nearly two weeks after the GRB.  
This is the strongest case yet for a GRB/supernova connection.  
It strengthens the case that a supernova also dominated the late afterglow of GRB 980326, and the case that GRB 980425 is related to SN 1998bw.
\end{abstract}


\section{Introduction}

The discovery of both an X-ray (Costa et al. 1997a) and optical (Groot et al. 1997) afterglow to GRB 970228 revolutionized the gamma-ray burst (GRB) field.  The mean temporal and spectral properties of this afterglow appeared to be consistent with the relativistic fireball model (Tavani 1997; Waxman 1997; Wijers, Rees, \& M\'esz\'aros 1997; Reichart 1997; Sahu et al. 1997; Katz \& Piran 1997), which predicted that GRBs would have afterglows at these and radio wavelengths (Paczy\'nski \& Rhoads 1993; Katz 1994; M\'esz\'aros \& Rees 1997).  However, now that the photometry has been finalized, and perhaps more importantly, now that nearly a dozen optical afterglows to GRBs are available for comparison, the afterglow of GRB 970228 is perhaps the most difficult of the observed afterglows to reconcile with the fireball model (see \S 2).  

The most problematic feature of this afterglow for the fireball model
is its extreme reddening with time:  V - I changes from $0.7 \pm 0.2$
mag 21 hours after the burst to $2.3 \pm 0.2$ mag 26 days after the
burst and $2.2 \pm 0.3$ mag 38 days after the burst.  Bloom et al.
(1999) found that the afterglow of GRB 980326 exhibits a similar
behavior, and argued that this is due to a supernova that overtook the
light curve of the afterglow about one week after the burst.

Unfortunately, late afterglow data is sparse in the case of GRB980326, and its redshift is not yet known.  Data is more plentiful, both temporally and spectrally, in the case of GRB 970228, and its redshift, $z = 0.695 \pm 0.002$ (Djorgovski et al. 1999), is known.  Therefore, in this Letter, we build on the groundbreaking work of Bloom et al. (1999), and investigate the suggestion of Dar (1999) that a supernova dominated the late afterglow of GRB 970228.
In \S 3, we argue that the R-band light curve of this afterglow favors this hypothesis.  In \S 4, we argue that the spectral energy distribution of this afterglow at late times also favors this hypothesis.  We draw conclusions in \S 5.

\section{Problems with the Relativistic Fireball Model Interpretation of the Afterglow of GRB 970228}

As mentioned in \S 1, the most problematic feature of the afterglow of GRB 970228 for the relativistic fireball model is the afterglow's extreme reddening with time.  The fireball model can accommodate mild reddening; e.g., the passage of either the synchrotron or cooling breaks of this model's spectrum (see e.g., Sari, Piran, \& Narayan 1998) through these bands would increase the V $-$ I color of the afterglow by $\sim 0.3$ mag.  However, in the case of GRB 970228, V $-$ I increased by $\approx$ 1.6 mag.

To better quantify how this afterglow changed from early to late times, we have collected from the literature the published optical and near-infrared (NIR) photometry of the afterglow, which we list in Table 1, as well as that of the underlying host galaxy, which we list in Table 2.  In Figure 1, we plot the BVRI photometry of the afterglow.  Excluding the 1.5-m Bologna University Telescope (BUT) photometry at $\approx$ 17 hours after the burst (see Guarnieri et al. 1997 for a discussion of these data), clear and distinct trends can be seen in the early ($t < 5$ days after the burst) and late ($t > 25$ days after the burst) subsets of these data:  the late afterglow appears to have faded more slowly than the early afterglow, and as already mentioned, the late afterglow appears to have been considerably redder than the early afterglow.  The intermediate afterglow (5 days $< t < 25$ days) appears to have been transitional between these two limiting behaviors.

To quantify these limiting behaviors, we have fitted the functional form $F_{\nu} = F_0\nu^at^b$ to these two subsets of the data.  For the early afterglow, we find that $a_{BVRI} = -0.61 \pm 0.32$ and $b = -1.58 \pm 0.28$ ($\chi^2 = 1.609$, $\nu = 2$); for the late afterglow, we find that $a_{VRI} = -4.31 \pm 0.30$ and $b = -0.89 \pm 0.12$ ($\chi^2 = 0.153$, $\nu = 3$).  In the case of the early afterglow, the afterglow dominated the host galaxy (see Tables 1 and 2); consequently, we could ignore the contribution of the host galaxy in this fit.  In the case of the late afterglow, all of the measurements, except for the Keck-II 10-m R-band measurement of Metzger et al. (1997b), were made from HST images; in these cases, the angular resolution was sufficient to separate the afterglow from the host galaxy.  In the case of the R-band measurement, we instead fitted the functional form $F_0\nu^at^b + F_R$, where $F_R$ is the flux density of the host galaxy in the R band.  Employing Bayesian inference, we adopted a prior probability distribution for $F_R$, given by the HST/STIS R-band measurement of Castander \& Lamb (1999b; see also Fruchter et al. 1999).  

To confirm that the afterglow is not consistent with a single power-law spectrum and a single power-law fading, we fitted the above functional forms to the entire BVRI data set, excluding the BUT 1.5-m photometry and the 2.5-m Isaac Newton Telescope (INT) B-band measurement at $\approx$ 10 days, which may be dominated by the host galaxy (no B-band measurement of the host galaxy is available for comparison).  We find that $\chi^2 = 78.487$ for $\nu = 11$ degrees of freedom.  If we restrict ourselves to the data set fitted to above, we find that $\chi^2 = 67.844$ for $\nu = 8$.  Comparing the model fitted above with this model, we find that $\Delta \chi^2 = 67.844 - 1.609 - 0.153 = 66.082$ for $\Delta \nu = 8 - 3 - 2 = 3$; using the $\Delta \chi^2$ test, this implies that the above ``two limiting power-law behaviors'' model is favored over this ``single power-law behavior'' model at the $10^{-13.5}$ confidence level.

The temporal index of the early afterglow, $b = -1.58 \pm 0.28$, is similar to that found by Costa et al. (1997b) in the 2 - 10 keV band ($b = -1.33^{+0.11}_{-0.13}$, 8 hours $\la t \la 4$ days), and to that found by Frontera et al. 1998 in the 0.1 - 2.4 keV band ($b = -1.50^{+0.35}_{-0.23}$, 8 hours $\la t \la 13$ days).  Furthermore, the spectral index of the early afterglow, $a_{BVRI} = -0.61 \pm 0.32$, is consistent with the optical to X-ray spectral index of the afterglow at these early times:  $a_{OX} \approx -0.7$.  Finally, both the spectral and temporal indices of the early afterglow are consistent (1) with the spectral and temporal indices of other afterglows, and (2) with the general expectations of the fireball model.  

However, the spectral index of the late afterglow, $a_{VRI} = -4.31 \pm 0.30$, is not consistent with the general expectations of the fireball model.  In principle, one could achieve this extremely negative spectral index if the afterglow were heavily extincted, either by our galaxy, by the host galaxy, or by the immediate vicinity of the burst (see e.g., Reichart 1998).  However, if that were the case, either the intrinsic spectral index of the early afterglow would have to be extremely positive, which would also defy the general expectations of the fireball model, or the level of extinction would have to increase with time, which is the opposite of what one expects:  the afterglow may destroy dust, but it will not create it.  Consequently, although the early afterglow appears to be consistent with the fireball model, the late afterglow does not.  Likewise, the late afterglow is equally unlikely to be due to a refreshed shock, which must also meet the general expectations of the fireball model.
However, the extremely negative spectral index of the late afterglow is consistent with the general expectations of a supernova at the redshift of this burst.  We return to this claim in \S 4.

\section{Evidence for a Supernova in the R-band Light Curve of the Afterglow of GRB 970228}

In this section we investigate whether the R-band light curve of the afterglow of GRB 970228 is consistent with a supernova overtaking the light curve of the early afterglow at late times; we investigate whether the spectral energy distribution of the late afterglow is consistent with this hypothesis in \S 4.  

First, we construct the R-band light curve.  We do this by scaling all non-R-band measurements to the R-band using the fitted functional forms of \S 2; we have again excluded the BUT 1.5-m photometry and the INT 2.5-m B-band measurement (see \S 2).  In the top panel of Figure 2, we plot the ground-based photometry of the afterglow plus the host galaxy, as well as the best-fit R-band light curve of the early afterglow from \S 2 plus the best-fit R-band flux density of the host galaxy from Table 2.  In the bottom panel of Figure 2, we plot the space-based photometry of the afterglow, as well as the best-fit R-band light curve of the early afterglow. 

Next, we construct an example supernova light curve.  We begin with the U-band light curve of the Type Ib/c supernova SN 1998bw, which may or may not be related to GRB 980425.\footnote{In addition to SN 1998bw, the fading X-ray source 1SAX J1935.3-5252 was found in the error circle of GRB 980425 (Pian et al. 1998a,b).  Since 1SAX J1935.3-5252 is typical of nearly every other GRB afterglow observed to date, and since SN 1998bw, and in particular, its low redshift of $z = 0.0085$ (Tinney et al. 1998), are not typical of any other GRB afterglow observed to date, the association between GRB 980425 and SN 1998bw is uncertain (see also Graziani, Lamb, \& Marion 1999).  However, for the purposes of this Letter, any Type Ib - Ic supernova light curve is sufficient.  We have chosen SN 1998bw primarily because it is well-sampled, and only secondarily because of its possible association with GRB 980425.}  
Galama et al. (1998) observed SN 1998bw in the U band between 9 and 56 days after this burst.  At earlier times we extrapolate to the U band from the B band light curve of Galama et al. (1998); at later times we extrapolate to the U band from the B band light curve of McKenzie \& Schaefer (1999). 
Next, we transform this U-band light curve to the redshift of GRB 970228, z = 0.695, which involves (1) shifting it to the R band, (2) stretching it in time, and (3) dimming it.  Here, we have set $\Omega_m = 0.3$ and $\Omega_{\Lambda} = 0.7$; other cosmologies yield similar results.  Finally, we have corrected the data for differences in Galactic extinction between the SN 1998bw line of sight in the U band ($A_U = 0.382$ mag), and the GRB 970228 afterglow line of sight in the R band ($A_R = 0.959$ mag), using the dust maps of Schlegel, Finkbeiner, \& Davis (1998)\footnote{Software and data available at http://astro.berkeley.edu/davis/dust/index.html.} in the case of the SN 1998bw line of sight, Castander \& Lamb (1999a; however, see also Fruchter et al. 1999) in the case of the GRB 970228 line of sight, and the Galactic extinction curve of Cardelli, Clayton, \& Mathis (1989) for $R_V = 3.1$; no attempt has been made to quantify the differences in extinction between these sources' host galaxies and immediate vicinities along their respective lines of sight.  This redshifted, Galactic extinction-corrected, example supernova light curve is also plotted in Figure 2.

Despite all of the uncertainties in this example supernova light curve, including the uncertainty in its peak luminosity, which can vary by factors of a few for Type Ib - Ic supernovae, we find the summed early afterglow plus host galaxy plus supernova light curve (Figure 2, top panel) and the summed early afterglow plus supernova light curve (Figure 2, bottom panel) to be remarkably consistent with the photometry.\footnote{The HST/STIS observation at 189 days after the burst was done in Clear Aperture mode (Fruchter et al. 1999); consequently, its measured value depends upon an assumption about the spectrum of the afterglow at this late time.  In Figures 1 and 2, we assume that the spectrum of the late afterglow did not evolve with time; if, for example, the late afterglow grew bluer with time, then Figures 1 and 2 overestimate the flux density of the afterglow at this late time.}

\section{Evidence for a Supernova in the Spectral Energy Distribution of the Late Afterglow of GRB 970228}

We have shown that the R-band light curve of the late afterglow of GRB 970228 is consistent with what one would expect from a supernova at the redshift of the burst.  In this section, we go a step further and show that the spectral energy distribution of the late afterglow is also consistent with that of a supernova at this redshift, and that this consistency spans five spectral bands.

First, we construct the spectral energy distribution of the late afterglow.
Between $\approx$ 30 and $\approx$ 38 days after the burst, Sahu et al. (1997) observed the afterglow in the V and I bands with HST/WFPC2, Metzger et al. (1997b) observed the afterglow plus host galaxy in the R band with the Keck-II 10-m, and Soifer et al. (1997) observed the afterglow plus host galaxy in the J and K bands with the Keck-I 10-m. 
From these R- and K-band measurements of the afterglow plus host galaxy, we subtract the contribution of the host galaxy (see Table 2); as no J-band measurement of the host galaxy is available, we replace this J-band measurement of the afterglow plus host galaxy with an upper limit.
Next, we scale these afterglow measurements to a common time, $\approx 38$ days after the burst, using the fitted functional form of the late afterglow from \S 2; as these measurements are nearly coincident in $\log{t}$, this adjustment is minor.  
Finally, we correct these afterglow measurements for Galactic extinction along the line of sight (see \S 3).  We plot the resulting Galactic extinction-corrected, five-band spectral energy distribution of the late afterglow in Figure 3.

For comparison, we also plot in Figure 3 the Galactic extinction-corrected, redshifted spectral energy distribution of SN 1998bw (see \S 3), which Galama et al. (1998) observed in the U, B, V, R, and I bands $22 \approx 38 (1 + z_{SN 1998bw}) / (1 + z_{GRB 970228})$ days after GRB 980425.  When transformed to the redshift of GRB 970228, these measurements span the R through J bands.  

To the limit of the uncertainties in the late afterglow spectral energy distribution measurements, and to the limit of the smaller wavelength range of the redshifted SN 1998bw spectral energy distribution, these distributions appear to be nearly identical.  Furthermore, the relativistic fireball model is not expected to produce a spectral break of this nature, especially so long after the burst.  Consequently, we conclude that the late afterglow of GRB 970228 was most likely dominated by a supernova.

\section{Conclusions}

In conclusion, while the early afterglow of GRB 970228 appears to have been consistent with the general expectations of the relativistic fireball model, the late afterglow does not appear to have been consistent with this model.  Nor was it similar to any of the other observed afterglows, with the sole exception of the afterglow of GRB 980326, which Bloom et al. (1999) attribute to a supernova that overtook the light curve.  We find that this was most likely the case with GRB 970228.  Not only is the light curve of the afterglow consistent with this hypothesis, but more convincingly, the spectral energy distribution of the late afterglow is also consistent with this hypothesis, and this consistency spans five spectral bands.  
Furthermore, the identification of the late afterglow of GRB 970228 with a supernova strengthens the case that other GRBs, in particular, GRB 980326 and GRB 980425, are indeed related to supernovae.  However, in the case of GRB 980425, many reservations remain (\S 3).

\acknowledgements

This research was supported in part by NASA grant NAG5-2868 and NASA contract NASW-4690.  D. E. R. thanks D. Q. Lamb for comments that greatly improved this Letter, as well as for his general enthusiasm for this research.

\clearpage

\begin{deluxetable}{ccccc}
\footnotesize
\tablecolumns{5}
\tablewidth{0pc}
\tablecaption{Optical and NIR Photometry of the Afterglow of GRB 970228\tablenotemark{a}}
\tablehead{\colhead{Band} & \colhead{Date\tablenotemark{b}} & \colhead{Magnitude} & \colhead{Telescope/Instrument} & \colhead{Reference\tablenotemark{c}}}
\startdata
B & 1997 Feb 28.86 & $22.4 \pm 0.4$ & BUT 1.5-m & 1 \nl
B & 1997 Mar 3.1 & $23.3 \pm 0.5$ & APO 3.5-m & 2 \nl
B & 1997 Mar 9.85 & $25.4 \pm 0.4$ & INT 2.5-m & 3 \nl
V & 1997 Feb 28.99 & $21.3 \pm 0.1$ & WHT 4.2-m & 3 \nl
V & 1997 Mar 4.86 & $> 24.2$ & NOT 2.5-m & 3 \nl
V & 1997 Mar 26.20 & $26.20^{+0.14}_{-0.13}$\tablenotemark{d} & HST/WFPC2 & 4 \nl
V & 1997 Apr 7.24 & $26.52^{+0.16}_{-0.18}$\tablenotemark{d} & HST/WFPC2 & 4 \nl 
V & 1997 Sep 4.71 & $28.10^{+0.24}_{-0.23}$\tablenotemark{d} & HST/STIS & 5 \nl
R & 1997 Feb 28.81 & $20.5 \pm 0.5$ & RAO 0.6-m & 6 \nl
R & 1997 Feb 28.83 & $21.6 \pm 0.3$ & BUT 1.5-m & 1 \nl
R & 1997 Mar 6.32 & $24.0$ & Keck-II 10-m & 7 \nl
R & 1997 Mar 9.90 & $24.0 \pm 0.2$ & INT 2.5-m & 3 \nl
R & 1997 Mar 13.00 & $24.3 \pm 0.2$ & NTT 3.5-m & 3 \nl
R & 1997 Apr 5.24 + 6.27 & $24.9 \pm 0.3$ & Keck-II 10-m & 8 \nl
I & 1997 Mar 1.00 & $20.6 \pm 0.1$ & WHT 4.2-m & 3 \nl
I & 1997 Mar 26.20 & $23.94^{+0.10}_{-0.09}$\tablenotemark{d} & HST/WFPC2 & 4 \nl
I & 1997 Apr 7.24 & $24.31^{+0.15}_{-0.11}$\tablenotemark{d} & HST/WFPC2 & 4 \nl
J & 1997 Mar 30.3 + 31.2 & $23.5 \pm 0.2$ & Keck-I 10-m & 9 \nl
H & 1998 Feb 24 & $> 25.9$\tablenotemark{d,e} & HST/NICMOS2 & 10 \nl
K & 1997 Mar 30.2 & $22.0 \pm 0.2$ & Keck-I 10-m & 9 \nl
\enddata
\tablenotetext{a}{A hyperliked version of this table can be found at http://astro.uchicago.edu/home/web/reichart/grb/grb.html.}
\tablenotetext{b}{UT in decimal days.}
\tablenotetext{c}{1. Guarnieri et al. 1997; 2. Margon et al. 1997; 3. van Paradijs et al. 1997; 4. Castander \& Lamb 1999b (see also Fruchter et al. 1999; Sahu et al. 1997); 5. Castander \& Lamb 1999b (see also Fruchter et al. 1999); 6. Pedichini et al. 1997; 7. Metzger et al. 1997a; 8. Metzger et al. 1997b; 9. Soifer et al. 1997; 10. Fruchter et al. 1999.}
\tablenotetext{d}{Afterglow magnitude; no contribution from the host galaxy.}
\tablenotetext{e}{3 $\sigma$.}
\end{deluxetable}

\clearpage

\begin{deluxetable}{ccccc}
\footnotesize
\tablecolumns{5}
\tablewidth{0pc}
\tablecaption{Optical and NIR Photometry of the Host Galaxy of GRB 970228\tablenotemark{a}}
\tablehead{\colhead{Band} & \colhead{Date\tablenotemark{b}} & \colhead{Magnitude} & \colhead{Telescope/Instrument} & \colhead{Reference\tablenotemark{c}}}
\startdata
V & 1997 Mar 26.20 + Apr 7.24 & $25.85^{+0.26}_{-0.23}$ & HST/WFPC2 & 1 \nl
R & 1997 Sep 4 & $25.5$ & Palomar 5-m & 2 \nl
R  & 1997 Sep 4.71 & $25.54^{+0.33}_{-0.22}$ & HST/STIS & 3 \nl
I & 1997 Mar 26.20 + Apr 7.24 & $24.84^{+0.46}_{-0.42}$ & HST/WFPC2 & 1 \nl
H & 1998 Feb 24 & $23.2 \pm 0.1$ & HST/NICMOS2 & 4 \nl
K & 1998 Apr 8 & $22.8 \pm 0.3$ & Keck-I 10-m & 4 \nl
\enddata
\tablenotetext{a}{A hyperliked version of this table can be found at http://astro.uchicago.edu/home/web/reichart/grb/grb.html.}
\tablenotetext{b}{UT in decimal days.}
\tablenotetext{c}{1. Castander \& Lamb 1999b (see also Fruchter et al. 1999; Sahu et al. 1997); 2. Djorgovski et al. 1997; 3. Castander \& Lamb 1999b (see also Fruchter et al. 1999); 4. Fruchter et al. 1999.}
\end{deluxetable}

\clearpage

\figcaption[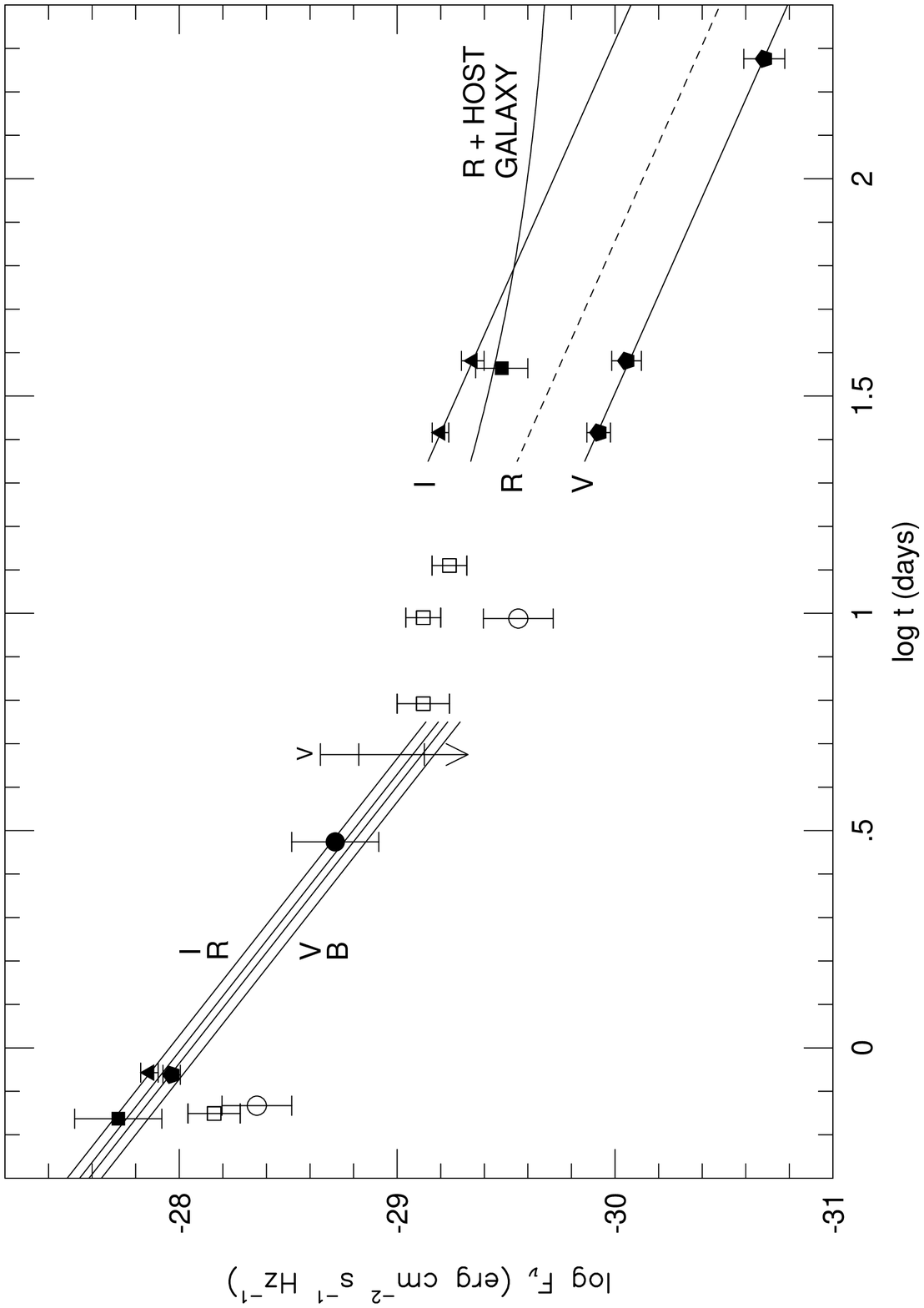]{The B-, V-, R-, and I-band light curves of the afterglow of GRB 970228 from the measurements of Table 1, and the best fits to the early and late subsets of these data (filled symbols, see \S 2).  Circles denote the B band, pentagons denote the V band, squares denote the R band, and triangles denote the I band.  The upper limit is 1, 2, and 3 $\sigma$.\label{228fig1.ps}}

\figcaption[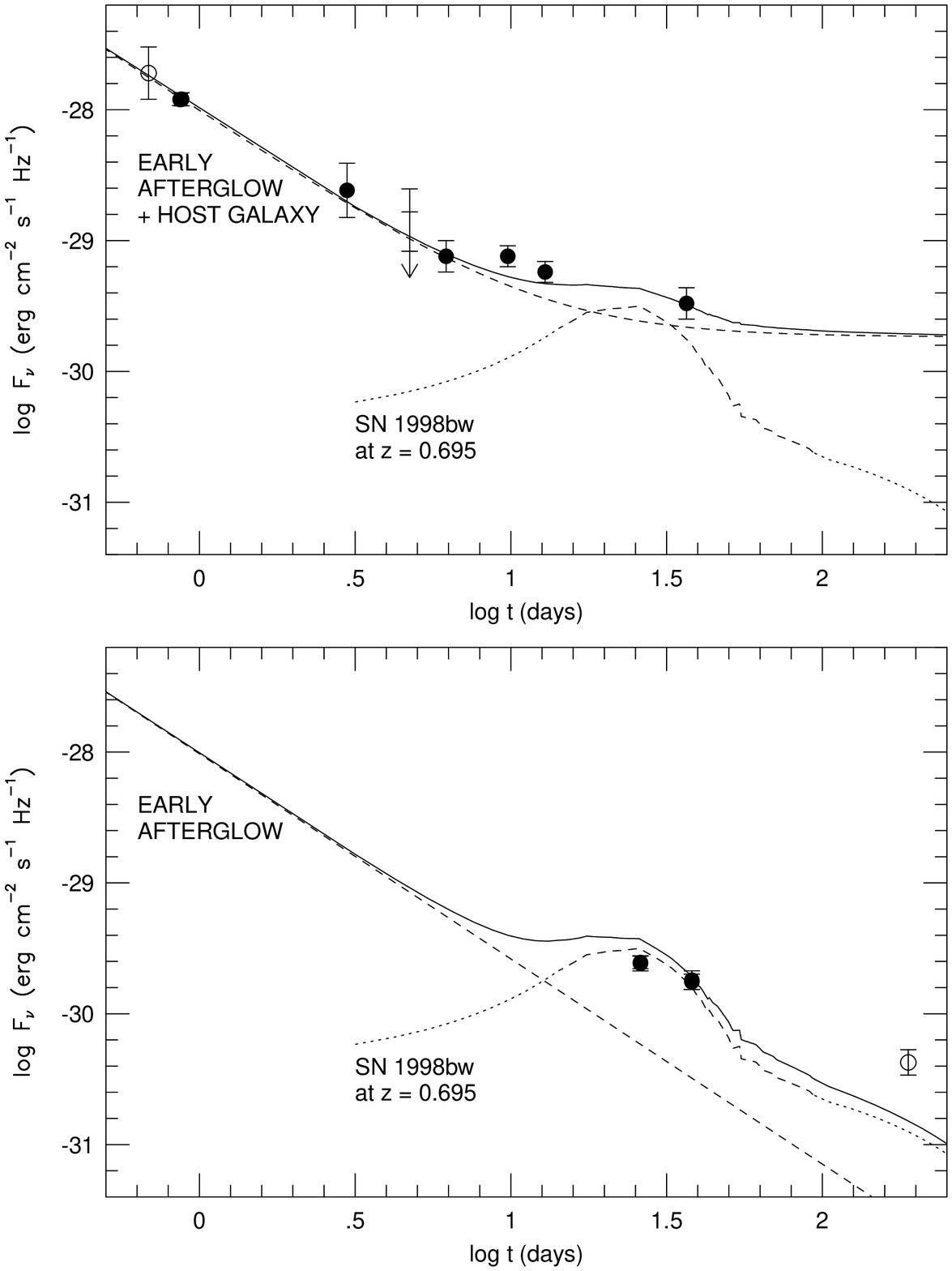]{Top panel:  The scaled (see \S 3), R-band light curve of the ground-based measurements of the afterglow plus host galaxy of GRB 970228, and the best-fit R-band light curve of the early afterglow from Figure 1 plus the best-fit R-band flux density of the host galaxy from Table 2.  Bottom panel:  The scaled, R-band light curve of the space-based measurements of the afterglow of GRB 970228, and the best-fit R-band light curve of the early afterglow from Figure 1.  Both panels:  To these best-fit light curves, we add the Galactic extinction-corrected, light curve of SN 1998bw, transformed to the redshift of GRB 970228 (see \S 3).  The dashed portion of this curve is derived from the U-band light curve of SN 1998bw; the dotted portions of this curve are derived by extrapolating to the U band from the B-band light curve of SN 1998bw (see \S 3).  
Unfilled circles denote unfiltered observations; these measurements may be in error by a color term (see \S 3 for a discussion of the HST/STIS observation at 189 days after the burst).\label{228fig2.ps}}

\figcaption[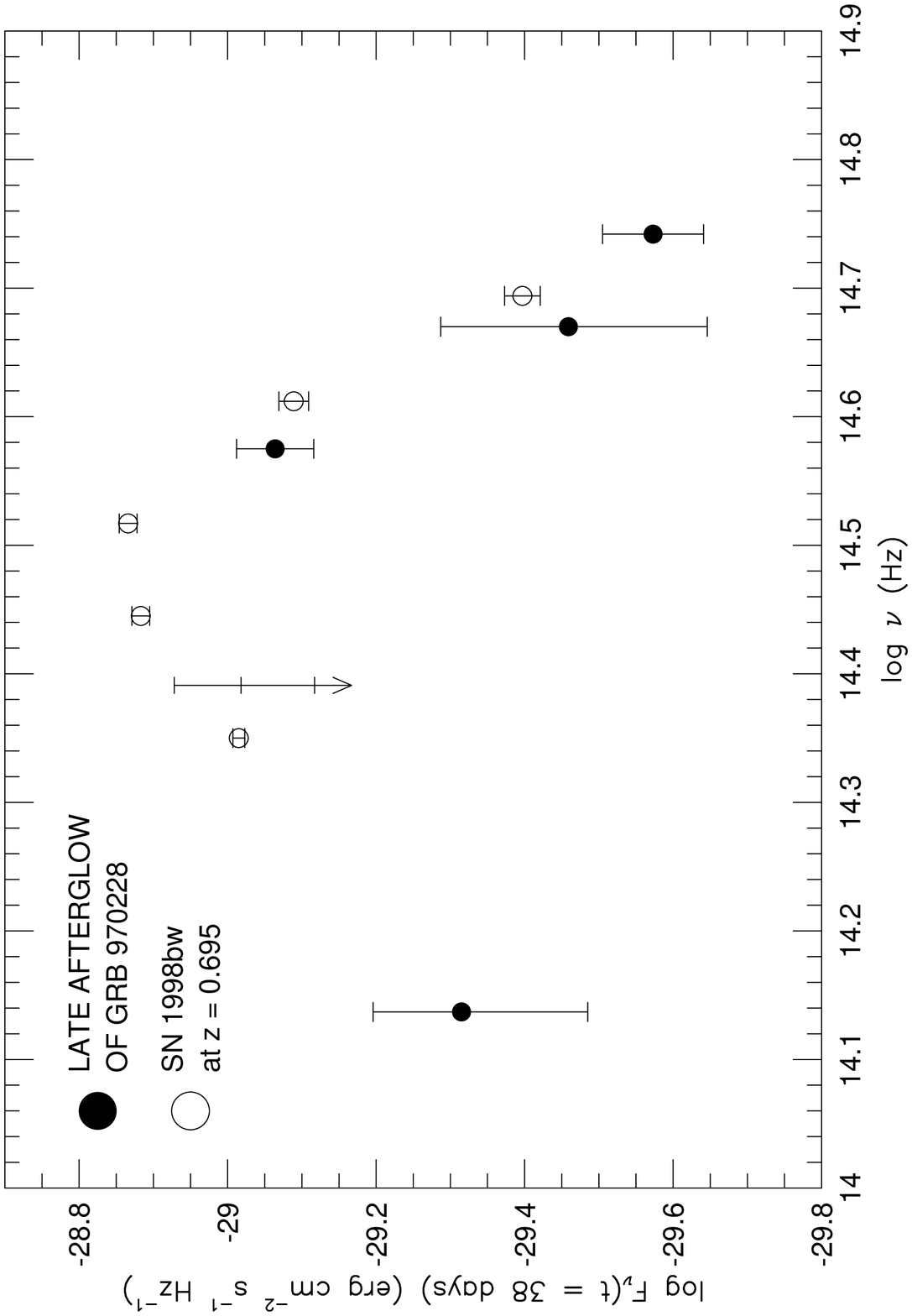]{The Galactic extinction-corrected spectral energy distributions of the late afterglow of GRB 970228 (filled circles), and of SN 1998bw (unfilled circles), transformed to the redshift of GRB 970228 (see \S 4).  The upper limit is 1, 2, and 3 $\sigma$.
\label{228fig3.ps}}

\clearpage

\setcounter{figure}{0}

\begin{figure}[tb]
\plotone{228fig1.ps}
\end{figure}

\clearpage

\begin{figure}[tb]
\plotone{228fig2.ps}
\end{figure}

\clearpage

\begin{figure}[tb]
\plotone{228fig3.ps}
\end{figure}


\clearpage

\begin{thebibliography}{}

\bibitem[Bloom et al. 1999]{bea99}
Bloom, J. S., et al. 1999, Nature, in press

\bibitem[Cardelli, Clayton, \& Mathis 1987]{ccd87}
Cardelli, J. A., Clayton, G. C., \& Mathis, J. S. 1987, ApJ, 345, 245

\bibitem[Castander \& Lamb 1999a]{cl99a}
Castander, F. J., \& Lamb, D. Q. 1999a, ApJ, in press

\bibitem[Castander \& Lamb 1999b]{cl99b}
Castander, F. J., \& Lamb, D. Q. 1999b, ApJ, in press

\bibitem[Costa et al. 1997a]{cea97a}
Costa, E., et al. 1997a, IAU Circular 6572

\bibitem[Costa et al. 1997b]{cea97b}
Costa, E., et al. 1997b, Nature, 387, 783

\bibitem[Dar 1999]{d99}
Dar, A. 1999, GCN Report 346

\bibitem[Djorgovski et al. 1997]{dea97}
Djorgovski, S. G., et al. 1997, IAU Circular 6732

\bibitem[Djorgovski et al. 1999]{dea99}
Djorgovski, S. G., et al. 1999, GCN Report 289

\bibitem[Fruchter et al. 1999]{fea99}
Fruchter, A. S., et al. 1999, ApJ, 516, 683

\bibitem[Frontera et al. 1998]{fea98}
Frontera, F., et al. 1998, A\&A, 334, L69

\bibitem[Galama et al. 1998]{gea98}
Galama, T. J., et al. 1998, Nature, 395, 670

\bibitem[Graziani, Lamb, \& Marion 1999]{glm99}
Graziani, C., Lamb, D. Q., \& Marion, G. H. 1999, ApJ, in press

\bibitem[Groot et al. 1997]{gea97}
Groot, P. J., et al. 1997, IAU Circular 6588

\bibitem[Guarnieri et al. 1997]{gea97}
Guarnieri, A., et al. 1997, A\&A, 328, L13

\bibitem[Katz 1994]{k94}
Katz, J. I. 1994, ApJ, 432, L107 

\bibitem[Katz \& Piran 1997]{kp97}
Katz, J. I., \& Piran, T. 1997, ApJ, 490, 772

\bibitem[Margon et al. 1997]{mea97}
Margon, B., et al. 1997, IAU Circular 6618

\bibitem[McKenzie \& Schaefer 1999]{ms99}
McKenzie, E. H., \& Schaefer, B. E. 1999, PASP, in press

\bibitem[M\'esz\'aros \& Rees 1997]{mr97}
M\'esz\'aros, P., \& Rees, M. J. 1997, ApJ, 476, 232

\bibitem[Metzger et al. 1997a]{mea97a}
Metzger, M. R., et al. 1997a, IAU Circular 6588

\bibitem[Metzger et al. 1997b]{mea97b}
Metzger, M. R., et al. 1997b, IAU Circular 6631

\bibitem[Paczy\'nski \& Rhoads 1993]{pr93}
Paczy\'nski, B., \& Rhoads, J. E. 1993, ApJ, 418, L5

\bibitem[Pedichini et al. 1997]{pea97}
Pedichini, F., et al. 1997, A\&A, 327, L36

\bibitem[Pian et al. 1998a]{pea98a}
Pian, E. 1998a, GCN Report 61

\bibitem[Pian et al. 1998b]{pea98b}
Pian, E. 1998b, GCN Report 69

\bibitem[Reichart 1997]{r97}
Reichart, D. E. 1997, ApJ, 485, L57

\bibitem[Reichart 1998]{r98a}
Reichart, D. E. 1998, ApJ, 495, L99

\bibitem[Sahu \etal 1997a]{sea97a}
Sahu, K. C., et al. 1997, Nature, 387, 476

\bibitem[Sari, Piran, \& Narayan 1998]{spn98}
Sari, R., Piran, T., \& Narayan, R. 1998, ApJ, 497, L17

\bibitem[Schlegel, Finkbeiner, \& Davis 1998]{sfd98}
Schlegel, D. J., Finkbeiner, D. P., \& Davis, M. 1998, ApJ, 500, 525

\bibitem[Soifer et al. 1997]{sea97}
Soifer, B., et al. 1997, IAU Circular 6619

\bibitem[Tavani 1997]{t97}
Tavani, M. 1997, ApJ, 483, L87

\bibitem[Tinney et al. 1998]{tea98}
Tinney, C., et al. 1998, IAU Circular 6896

\bibitem[van Paradijs et al. 1997]{vea97}
van Paradijs, J., et al. 1997, Nature, 386, 686

\bibitem[Waxman 1997]{w97}
Waxman, E. 1997, ApJ, 485, L5

\bibitem[Wijers, Rees, \& M\'esz\'aros 1997]{wrm97}
Wijers, R. A. M. J., Rees, M. J., \& M\'esz\'aros, P. 1997, MNRAS, 288, L51

\end{thebibliography}
\end{document}